\documentclass[12pt,titlepage]{article}
\usepackage{setspace}
\usepackage{epsfig}
\usepackage{ae}
\usepackage{aecompl}
\usepackage{amsmath}
\usepackage{graphicx}
\usepackage[round,numbers,sort&compress]{natbib}
\begin{document}

\title{Field Theoretic Study of Bilayer Membrane Fusion III: Membranes 
with Leaves of Different Composition}

\author{J.-Y. Lee and M. Schick\\
Department of Physics\\ University of Washington,  Box
  351560, Seattle, WA 98195-1560}

\date{} 
\maketitle

\begin{abstract} 
We extend previous work on homogeneous bilayers to 
calculate the barriers to fusion of planar bilayers 
which contain two different amphiphiles, 
a lamellae-former and a hexagonal former, with different compositions of the two
in each leaf. Self-consistent field theory is employed, and both standard and 
alternative pathways are explored. We first calculate these barriers
as the amount of hexagonal former is increased equally in both leaves
to levels appropriate to the 
plasma membrane of human red blood cells. We follow these barriers  
as the composition 
of hexagonal-formers is then increased in the cis layer and decreased in the trans layer, again
to an extent comparable to the biological system. We find that, while the fusion pathway
exhibits two barriers in both the standard and alternative pathways, 
in both cases the magnitudes of these barriers are comparable to one another,
and small, on the order of 13 $k_BT$. As a consequence, one expects that
once the bilayers are brought sufficiently close to one another to 
initiate the process, fusion should occur rapidly.

\end{abstract}

\section{Introduction}
In spite of the importance of membrane fusion to biological processes
such as endocytosis, intra-cell trafficking, and viral infection, and
in spite of the increased attention devoted to it,
the process is still not well understood. In particular, it is unclear
what the sequence of events along the path to fusion is, which of those
events presents the greatest barrier to fusion, and what the magnitude
of that barrier is. 

\begin{figure}
   \begin{center}
       \includegraphics*[width=5.25in]{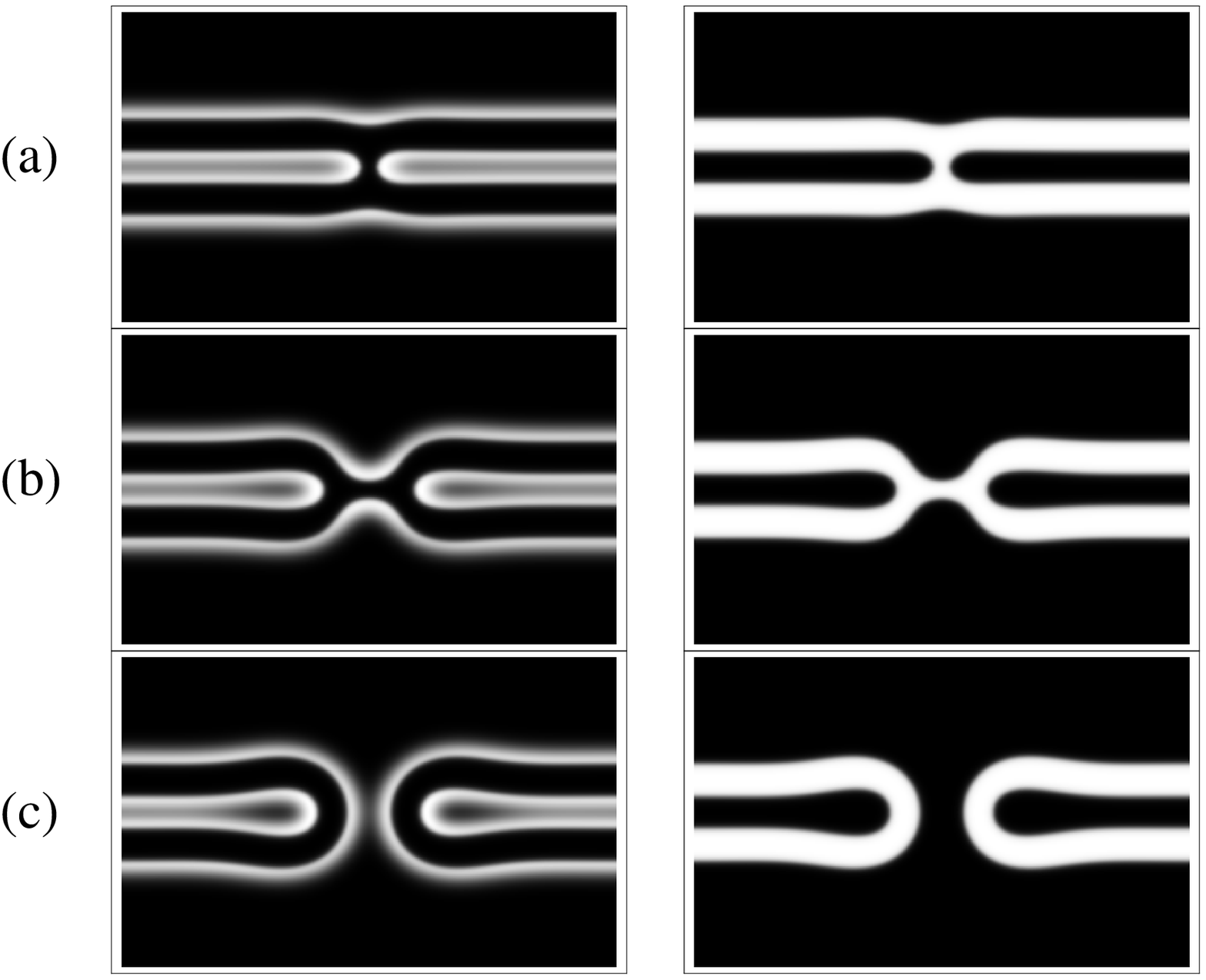}
      \caption{}
      \label{fig1}
   \end{center}
\end{figure}

The initial stages of the sequence are relatively clear 
\cite{Chernomordik03,Cohen04}. The membranes to be fused must be brought 
sufficiently close to one another, within a few nanometers. In order to do 
so, water must be removed, which takes energy. Presumably this is provided 
by fusion proteins in biological systems, but can, in laboratory ones in 
which such proteins are absent, be provided simply by ordinary depletion 
forces \cite{Evans02}. As a result of the decrease of water, the free 
energy per unit area of the system increases; in other words, the system 
is now under tension. The free energy can be reduced if the system sheds 
area. Fusion, which accomplishes this, is one possible response of the 
system to that tension. The next stage in the process is that, locally, 
some lipid tails in the membrane leaves which are closest to one another, 
the cis leaves, flip over and embed themselves in the hydrophobic 
environment of the cis leaf of the other bilayer, thereby forming a 
``stalk'' \cite{Kozlov83}, as depicted in Fig. \ref{fig1}(a). This
process is consistent with  
experimental evidence (\cite{Chernomordik03} and references therein), 
and has been seen directly in 
simulations of coarse-grained, microscopic, models of membranes 
\cite{Mueller02,Mueller03,Stevens03,Marrink03,Smeijers06}.

The next stage is unclear, and several possibilities have been proposed. 
The original suggestion \cite{Kozlov83} was that the stalk expands 
radially from an axis perpendicular to the bilayers, as in Fig. \ref{fig1}(b). The cis layers 
retract leaving a hemifusion diaphragm which consists only of the leaves 
of the two membranes which were initially furthest from one another, the 
trans leaves. Note that membrane area has been reduced as the hemifusion 
diaphragm now consists only of two, trans, leaves in place of the original 
four, two cis and two trans. The appearance of a hole in this hemifusion 
diaphragm completes the formation of the fusion pore, Fig. \ref{fig1}(c). On the basis of 
phenomenological modeling similar to that employed earlier 
\cite{Kozlov83}, a second scenario was suggested: that the pore forms 
without significant radial expansion of the stalk 
\cite{Siegel93,Kuzmin01}. A third possibility was revealed by simulations 
of coarse-grained, microscopic models 
\cite{Noguchi01,Mueller02,Mueller03}. In this, which we denote the first
stalk-hole 
mechanism, the stalk does not expand radially, but elongates 
asymmetrically. Its presence makes more favorable the formation nearby of 
a hole in either bilayer by reducing the line tension of the hole 
\cite{Katsov06}. The stalk then surrounds the hole which also produces a hemifusion 
diaphragm, as in the standard stalk mechanism, but one which consists here 
of a cis and trans layer of one of the original 
bilayers. The appearance of a second hole, this in the hemifusion 
diaphragm, then completes the fusion pore. A hemifusion diaphragm is also consistent with experimental evidence (\cite{Lentz06} and references therein.) In a variant of this mechanism, 
denoted the second stalk-hole mechanism, 
the second hole appears before the first is surrounded. The mobile stalk 
then surrounds them both forming the fusion pore. 
After 
formation of the fusion pore, the pore expands to further eliminate area 
and thus reduce the system's free energy.
Simulations of 
coarse-grained, microscopic, models have observed the original mechanism 
\cite{Noguchi01,Marrink03,Smeijers06}, and also the stalk-hole mechanism 
\cite{Noguchi01,Mueller02,Mueller03,Marrink03,Stevens03,Smeijers06}. 
If the path to fusion is not well established, neither is the limiting 
free energy barrier of the process. It had been thought, on the basis of 
phenomenological calculations, that the free energy to form the initial 
stalk was so large, that its formation could well be the barrier to 
fusion. Improvements in the way the stalk was modeled \cite{Markin02}, and 
in the phenomenological free energy describing the elastic properties of 
the membrane \cite{Kozlovsky02} which forms the stalk, resulted in a 
marked reduction in the estimate of the free energy of formation of the 
stalk. For a bilayer with symmetric leaves characterized by a spontaneous 
curvature appropriate to dioleoylphosphatidylcholine (DOPC), this quantity 
is estimated by Kozlovsky and Kozlov \cite{Kozlovsky02} to be 43$k_BT$, 
and by Kuzmin et al. \cite{Kuzmin01} to be about 25$k_BT$. In contrast to 
phenomenological theories, self-consistent field theory has been applied 
to a coarse-grained microscopic model of a symmetric membrane 
\cite{Katsov04}, resulting in an even lower estimate of 13$k_BT$. 
Irrespective of the particular number, it would not appear that the formation 
of the stalk presents the largest barrier to fusion.

If stalk formation is not the rate limiting process in fusion, what is? In 
the standard picture in which the stalk expands radially into a hemifusion 
diaphragm, it is the formation of this structure which takes a great deal 
of energy. For a symmetric bilayer of DOPC, a diaphragm of modest radius 
of 2.5nm costs on the order of 80$k_BT$, if one uses the estimate of 
Kozlovsky and Kozlov \cite{Kozlovsky02} for the diaphragm line tension. 
How large the diaphragm must become before a pore forms is not clear from 
this calculation. Kuzmin et al. \cite{Kuzmin01} consider a modified stalk 
and a different radial symmetric intermediate, a pre-pore. They find its 
energy, about 60 $k_BT$, to be less than that of a hemifusion diaphragm, 
and the largest along the fusion pathway. Self-consistent field 
calculations examined both the classical pathway \cite{Katsov04} and the 
first stalk-hole mechanism \cite{Katsov06}, and located the barriers to fusion 
for symmetric bilayers. In the former, the largest barrier occurred when 
the hemifusion diaphragm expanded to a radius which was of the same order 
of the hydrophobic thickness of a bilayer. Pore formation followed. The 
value of the barrier ranged from about 25 to 65 $k_BT$, depending upon the 
tension and the architecture of the amphiphiles. The barrier decreases 
with increasing tension and as the architecture tends toward  
dioleoylphosphatidylethanolamine, 
DOPE, and away from DOPC. Calculated barriers in the stalk-hole mechanism tended to be 
somewhat smaller than in the standard mechanism, but only by a few $k_BT$. 
Thus the two mechanisms seem to be comparable in terms of their 
energetics, at least for the symmetric membranes examined.

Biological membranes are not symmetric, however. In human red blood cell 
membranes, for example, most of the cholinephospholipids, sphingomyelin 
(SM) and phosphatidylcholine (PC), are found in the outer, ectoplasmic, 
leaf, and most of the aminophospholipids, phosphatidylethanolamine (PE) 
and phosphatidylserine (PS), are found in the inner, cytoplasmic, leaf 
\cite{Rothman77,Devaux91}. In particular the mole percent of PC in the 
outer/inner leaf is 22/8, of SM is 20/5, of PS is 0/10, and of PE is 8/27 
\cite{Rothman77}. To maintain this imbalance costs energy 
\cite{Seigneuret84}, therefore it is reasonable to assume that it plays 
some physiological function. One suggestion is that this imbalance 
promotes fusion in intracellular events \cite{Eastman92,Bailey94,Chernomordik95}. The reasoning 
is as follows. Of the four major lipid groups cited above, three of them, 
SM, PC, and PS \cite{Hope80} form bilayers under physiological conditions. 
They make up 65\% of the total bilayer, but 84\% of the outer, trans, 
leaf. PE, however, does not form lamellae, but rather an inverted 
hexagonal phase \cite{Gruner83}.  It has often been noted \cite{Siegel86} that 
regions of this non-bilayer phase resemble the non-bilayer configurations 
posited to occur in fusion. Furthermore, PE resides predominantly in the 
inner, cytoplasmic, leaf of the plasma membrane. While it makes up only 
35\% of the total bilayer composition of human red blood cell membranes,
it comprises 54\% of the inner 
leaf. It is presumed to also reside predominantly in the outer leaf of a 
bilayer vesicle within the cell, as the outer leaf of such a vesicle would 
make contact with the inner leaf of the plasma membrane during fusion of 
the vesicle and plasma membrane, and thereby have the opportunity to 
exchange lipid content. But it is precisely the inner leaf of the plasma 
membrane and the outer leaf of a vesicle which would be closest to one 
another during fusion (i.e. would be the cis leaves), and would undergo 
the largest deviation from a planar configuration. Hence the enhanced 
concentration of PE in these leaves would presumably promote fusion.

There is much experimental evidence to support the view that the presence 
of hexagonal-forming lipids in the cis leaves enhances fusion. In 
particular, model membranes (i.e. which have equal composition in both 
leaves) fuse readily when composed of a mixture of PE and PS approximating that of the inner 
leaf of the erythrocyte membrane \cite{Hope83}, while those 
consisting of PC and SM do not. Asymmetric membranes were investigated by 
Eastman et al. \cite{Eastman92} who utilized dioleoylphosphatidic acid 
(DOPA), a lipid with a head group smaller even than  PE, which they could 
move from cis to trans layers by applications of a pH gradient. When DOPA 
was present in the cis layers in modest amounts, 5 mol percent, fusion of 
large unilamellar vesicles occurred readily on the addition of Ca$^{2+}$. 
However when DOPA was sequestered in the trans monolayer, little or no 
fusion was observed. Conversely, if one adds to the cis layer lauroyl 
lysophosphatidylcholine, which has a large head group when compared to its 
single tail, fusion is inhibited dramatically \cite{Chernomordik95}.

As important as this asymmetry appears to be to the process of fusion, it 
is little addressed in theoretical calculations. In phenomenological ones, 
it has been accounted for by allowing the inner and outer leaves to be 
characterized by different spontaneous curvatures. In this way Kozlovsky 
and Kozlov \cite{Kozlovsky02} predict that the free energy of stalk 
formation depends essentially only on the spontaneous curvature of the cis 
leaves, and decreases rapidly as this curvature is made more negative, 
(i.e. as one proceeds from the lamellae-formers towards the
hexagonal -formers). A similar calculation and result follows for the free energy of 
formation of the hemifusion diaphragm \cite {Kozlovsky002}.  There are no 
direct results for the effect of the asymmetry on the largest barrier to 
fusion. However, by treating the entire cis layer as having the same 
spontaneous curvature, the calculation cannot capture the effect 
that hexagonal forming lipids can 
respond {\em locally} to an environment in which the leaves 
are locally deformed 
i.e. where their distribution, in general, is not uniform \cite{Li00}. 
Simulations of fusion have not yet considered the effects of asymmetry,
presumably because the asymmetric distribution represents a constrained 
equilibrium, a situation more difficult to handle than an unconstrained one.

In this paper, we extend the application of self-consistent field theory 
to microscopic models of membranes \cite{Katsov04,Katsov06} and consider 
two important effect noted above; that the bilayer leaves consist of at 
least two classes of lipids, lamellae-formers and hexagonal-formers, and 
that these lipids are distributed asymmetrically with respect to the cis 
and trans layers.  We shall deal with  these effects in two stages. 
First we consider the 
effect on the barriers to fusion due to the presence of 
two kinds of amphiphiles 
in leaves of identical composition, as in artificial 
membranes. We do this for the standard mechanism, and for both the first
and second stalk-hole fusion 
mechanisms. We find that the barriers are reduced appreciably because the 
hexagonal-forming amphiphiles can go to the regions where they relieve the 
most strain \cite{Li00}. The barriers in the two variants of the
stalk-hole mechanism are not very different from one another. 
 We then consider the same overall composition, 
but redistribute the two amphiphiles asymmetrically, with the 
hexagonal-formers being more concentrated in the cis leaves. The
barriers in the standard fusion mechanism and in the second stalk-hole
mechanism
are calculated.
The overall effect of having two such different amphiphiles distributed 
unequally between the two leaves is dramatic. 
The major 
barriers to fusion in the two scenarios are reduced to such an extent
that they are now  
comparable to the rather small initial barrier to stalk 
formation. This barrier is not affected appreciably by 
the addition of the hexagonal-formers nor by their asymmetric distribution, 
and remains on the order of 13 $k_BT$. As a result, the fusion pathway consists of
two small barriers. Once bilayers are brought sufficiently close to initiate the process, 
fusion should therefore proceed rapidly. 

\section{The model}
 The model is similar to that employed earlier \cite{Katsov04,Katsov06},
so we will only discuss here the necessary extensions. We consider a
system of two different amphiphiles, AB block copolymers, denoted 1 and
2, and solvent of A homopolymer. The volumes occupied by a solvent chain of
$N$ segments, and of a chain of amphiphile 1, also taken to be of $N$
segments, are $Nv$, where $v$ is the volume of each segment. The
volume occupied by a chain of amphiphile 2  of $\tilde{\alpha} N$ segments is 
$\tilde{\alpha} Nv$.
The fraction of hydrophilic, A, monomers in amphiphile 1 is $f_1$ and
that in amphiphile 2 is $f_2$. In our subsequent calculations we shall
take $f_1=0.4$, close to the value of 0.43 which would characterize
DOPC, and $f_2=0.294$, approximately 
the value characterizing DOPE \cite{Katsov04}. In
order that the hydrophobic length of the two different amphiphiles be
the same, we require $(1-f_1)Nv=(1-f_2)\tilde{\alpha} Nv$ so that 
${\tilde\alpha} =0.85.$
Thus we have two amphiphiles with the same hydrophobic length, but
different hydrophilic lengths. Amphiphile 2 is a hexagonal-former with a
smaller hydrophilic head group than amphiphile 1, which is a 
lamellae-former. We denote the local volume fraction of hydrophilic elements of
amphiphile 1 to be $\phi_{A,1}({\bf r})$, of amphiphile 2 to be 
$\phi_{A,2}({\bf r})$, and
of the solvent to be $\phi_{A,s}({\bf r})$. The total local volume
fraction of hydrophilic elements is denoted 
\begin{equation}
\phi_A({\bf r})=\phi_{A,1}({\bf r})+\phi_{A,2}({\bf r})+\phi_{A,s}({\bf
r}).
\end{equation}
Similarly the total local volume fraction of hydrophobic elements is
\begin{equation}
\phi_B({\bf r})=\phi_{B,1}({\bf r})+\phi_{B,2}({\bf r}).
\end{equation}
The 
amounts of each of the components are controlled by activities,
$z_1$, $z_2$, and $z_s$. The system is taken to be incompressible and of volume $V$. 
Because of the incompressibility constraint, only two of
the activities are independent. 
Within the self-consistent field
approximation, the excess free energy, $\delta\Omega^{sym}(T,A,z_1,z_2,z_s)$, 
of the bilayer system of area $A$,
is given by
\begin{eqnarray}
\label{free1}
\frac{Nv}{k_BT}\delta\Omega^{sym}&=&-z_1Q_1-z_2Q_2-z_sQ_s\nonumber \\
                  &&+\int d{\bf r}[\chi N\phi_A({\bf r})\phi_B({\bf r})-
w_A({\bf r})\phi_A({\bf r})-w_B({\bf r})\phi_B({\bf r})\nonumber \\
                  &&-\xi({\bf r})(1-\phi_A({\bf r})-\phi_B({\bf r}))],
\end{eqnarray} 
where $Q_1(T,[w_A,w_B])$, $Q_2(T,[w_A,w_B]),$ and $Q_s(T,[w_A])$ are the
configurational parts of the single chain partition functions of amphiphiles 1 and 2 and of
solvent. They have the dimensions of volume, and 
are functions of the temperature, $T$, which is inversely
related to the Flory interaction $\chi$, and functionals of
the fields $w_A$ and $w_B$. 
These fields, and the Lagrange multiplier
$\xi({\bf r})$, which enforces the local incompressibility condition, are
determined by the self-consistent equations
\begin{eqnarray}
\label{sceq1}
w_A({\bf r})&=&\chi N\phi_B({\bf r})+\xi({\bf r}),\\
\label{sceq2}
w_B({\bf r})&=&\chi N\phi_A({\bf r})+\xi({\bf r}),\\
1&=&\phi_A({\bf r})+\phi_B({\bf r}),\\
\phi_A({\bf r})&=&-z_1\frac{\delta Q_1[w_A,w_B]}{\delta w_A({\bf r})}-
z_2\frac{\delta Q_2[w_A,w_B]}{\delta w_A({\bf r})}-
z_s\frac{\delta Q_s[w_A]}{\delta w_A({\bf r})},\\
\label{sceq5}
\phi_B({\bf r})&=&-z_1\frac{\delta Q_1[w_A,w_B]}{\delta w_B({\bf r})}-
z_2\frac{\delta Q_2[w_A,w_B]}{\delta w_B({\bf r})}.
\end{eqnarray}
The partition functions are obtained from the solution of a modified 
diffusion equation, as detailed as in the first paper in this series 
\cite{Katsov04}, and the barriers to fusion are calculated for the 
standard and for the stalk/hole mechanisms as in the previous two papers 
\cite{Katsov04,Katsov06}. The free energy in the self consistent field 
approximation , $\delta\Omega_{scf}^{sym}$, is obtained by inserting into the free energy of
Eq. \ref{free1} the functions which satisfy 
the self-consistent equations \ref{sceq1}-\ref{sceq5} with the result 
 \begin{eqnarray}
 \label{scsymfree}
  \frac{Nv}{k_BT}\delta\Omega^{sym}_{scf}(T,A,z_1,z_2,z_s)&=&-z_1Q_1(T,[w_A,w_B])-z_2Q_2(T,[w_A,w_B])\nonumber\\
  &-&z_sQ_s(T,[w_A])-\int\ d{\bf r}\chi N\phi_A({\bf r})\phi_B({\bf r}),
\end{eqnarray}
where we have set $\int \xi({\bf r})d{\bf r}=0$.

Calculation of the barrier to fusion in the standard mechanism is
relatively straightforward because all intermediates, the stalk, hemifusion
diaphragm, and pore, are characterized by axial symmetry about the $z$
axis, and reflection symmetry in the $xy$ plane. The former symmetry is
absent in the intermediates of the stalk-hole mechanisms. In order to
make tractable the calculation of the barrier along this path, the
actual intermediates were approximated by intermediates constructed from
segments of configurations which possessed both symmetries and whose
free energy, therefore, were
easily obtained \cite{Katsov06}.

\begin{figure}
   \begin{center}
   \includegraphics*[width=5.25in]{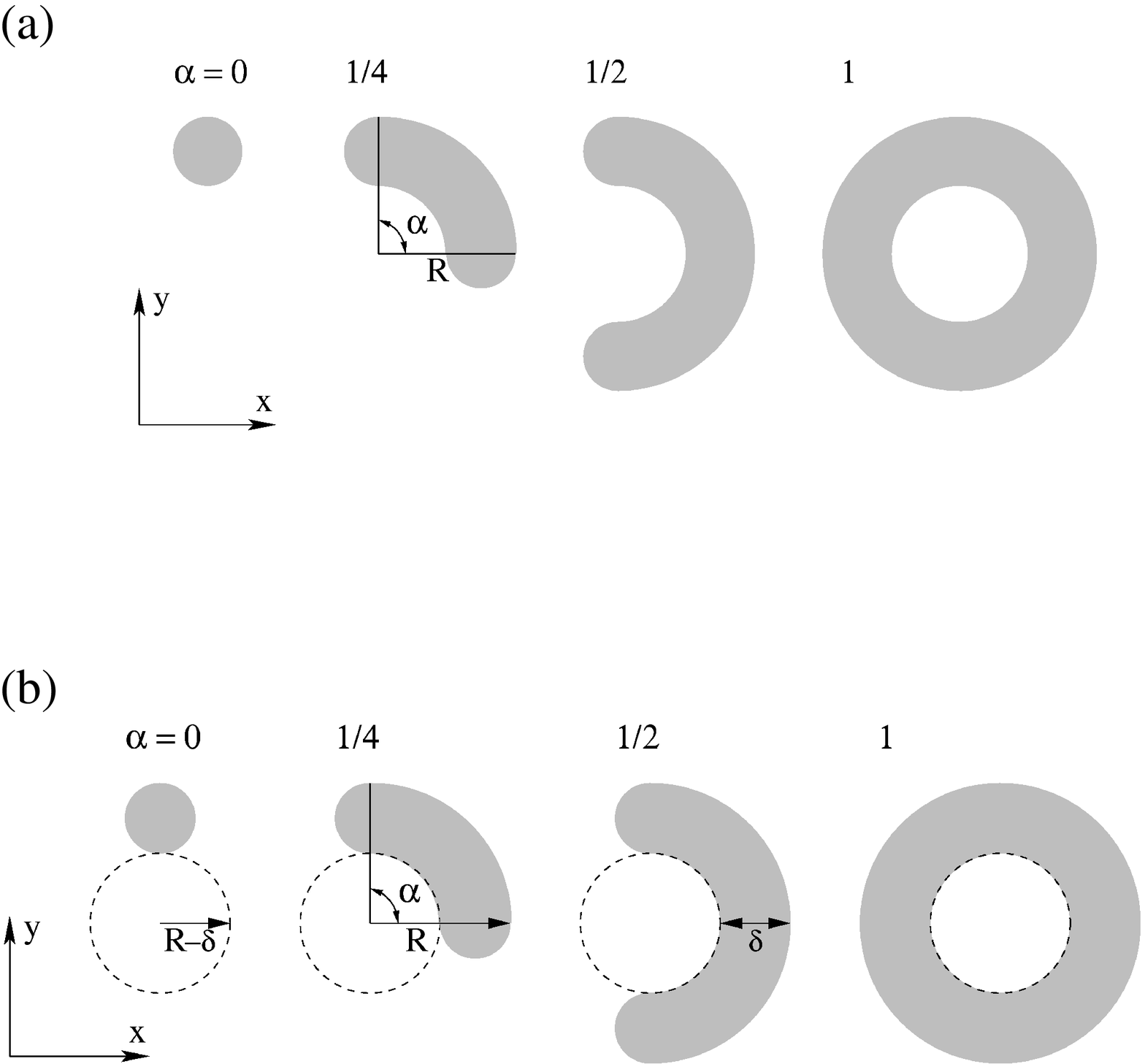}
      \caption{}
      \label{complex}
   \end{center}
\end{figure}

Just before formation of the stalk-hole complex, the elongated stalk was
treated as if it were in the shape of a circular arc with a fractional
angle,
$0\leq\alpha\leq 1$, and radius $R$, as shown schematically in
Fig. \ref{complex}(a). Its free energy is 
\begin{equation}
F_1(R,\alpha)=\alpha F_{IMI}(R)+F_s,
\label{freebefore}
\end{equation}
where $F_{IMI}$ is the energy of the structure shown at the extreme
right of Fig. \ref{complex}(a) which corresponds to $\alpha=1$, and
$F_s$ is the free energy of a stalk. This is  because it is the 
sum of the energies of the two end caps of a structure for
which $\alpha\neq 1$, and these two end caps together make a stalk.

Just after formation of the stalk-hole complex in the first stalk-hole
mechanism, there is a hole in one of the two bilayers
\cite{Loison04,Tolpekina04,Tieleman03,Leontiadou04,Groot01}.
which is partially
surrounded by the elongated stalk. This intermediate
is approximated by the configuration shown in Fig. \ref{complex}(b)
whose free energy is
\begin{equation}
F_2(R,\alpha)=\alpha F_{HI}(R)+(1-\alpha)F_H(R-\delta)+F_d.
\label{freeafter1}
\end{equation}
Here $F_{HI}$ is the free energy of the structure with $\alpha=1$ in
which the stalk would have completely surrounded the hole forming a
hemifusion intermediate, $F_H(R-\delta)$ is the free energy of a hole 
of radius $R-\delta$ in a bilayer, and $F_d$ is the free energy of the
defects at the end of the arc. Equality of the free energies of Eqs. \ref{freebefore}  
and \ref{freeafter1} defines a ridge line in the space of parameters
$\alpha$ and $R$, and the minimum of this ridge defines a saddle point
along this fusion path.

In the second stalk-hole mechanism,  just after formation of
the stalk-hole complex, there are two holes, one in each bilayer,
partially surrounded by the elongated stalk. Again the picture is as in
Fig. \ref{complex}(b), but now the circular object in the center of the
figure represents the {\em two} holes, rather than the one as
previously. Thus the figure at the extreme right now represents a fusion
pore. The free energy of this configuration is
\begin{equation}
F_3(R,\alpha)=\alpha F_{pore}(R)+(1-\alpha) F_{2H}(R-\delta)+F_d^{\prime}.
\label{freeafter2}
\end{equation}
Here $F_{pore}(R)$ is the free energy of a pore of radius $R$,
$F_{2H}(R-\delta)=2F_H(R-\delta)$ is the free energy of two holes, each
of radius $R-\delta$, one
above the other, and $F_d^{\prime}$ the energy of the two defects at the
end of the arc. Again equality of Eqs. \ref{freebefore} and
\ref{freeafter2} defines a ridge line in the space of parameters
$\alpha$ and $R$. The minimum along this ridge defines the fusion
barrier along this second stalk-hole pathway.

\section{Results for symmetric bilayers}
We first show in Fig. \ref{oldmech} 
how the addition of the hexagonal-forming amphiphiles 
affect the barrier to fusion in the standard mechanism.  
We plot there in solid lines the free energy of the stalk which expands into a 
hemifusion diaphragm  
as a function of the structure's radius divided by the radius of gyration, $R_g$, of 
the larger amphiphile.  (The hydrophobic thickness of a single bilayer 
composed of amphiphiles with $f=0.4$ is 
$2.7\ R_g$.) When the radius is smaller than about 0.5 $R_g$, we find no
stable stalk solution of the self-consistent equations.  We have taken the volume
$Nv$ which appears in the free energy, Eq. \ref{scsymfree}, to be $Nv=1.54 R_g^3$, 
as in our previous work \cite{Katsov04,Katsov06}. 
The four solid curves in Fig. \ref{oldmech} correspond to volume fractions of the 
hexagonal-former of 0, 0.04, 0.11, and 0.17 from top to bottom. The 
dotted curves show the free energies of fusion pores for the same 
volume fractions. We take the barrier to fusion to be that value at which 
the free energies of a hemifusion diaphragm and fusion pore of the same 
radius are equal. The bilayer is under a tension of 
$\gamma/\gamma_0=0.2$, where $\gamma_0$ is the interfacial 
free energy per unit area between coexisting solutions of hydrophobic and 
hydrophilic homopolymers at the same temperature. At larger values of the 
radius of the hemifusion diaphragm than shown in the figure, the free 
energy of the diaphragm decreases due to the tension. 
One sees from Fig. \ref{oldmech} that the barrier to fusion does indeed
decrease with the addition of hexagonal-formers. As can be seen in the
figure, this reduction comes about both because of the reduction in 
energy of the fusion pore and of the hemifusion diaphragm. 
The reduction in the pore energy is due to the effect of the hexagonal-formers
which can go to the sharp bend of the cis leaf existing in the pore.
Similarly the reduction in the energy of the hemifusion diaphragm is due
to the hexagonal-formers concentrating at the rim of the diaphragm.
This is shown in
Fig. \ref{nonuniform}. In (a) the volume fractions of the heads, dashed line, and tails,
solid line, of 
the hexagonal-forming amphiphile far from the hemifusion 
diaphragm are shown as a function of $z/R_g$.  In (b) we show the volume
fractions of the 
hexagonal-forming amphiphile in a cut through the hemifusion diaphragm itself in
the plane of reflection symmetry, the $z=0$ plane, as a function of the
radial coordinate, $\rho/R_g$.
(Recall that such a hemifusion diaphragm is shown in Fig. \ref{fig1}(b).) 
One sees that
the diaphragm has an approximate radius of  $5R_g.$ A comparison 
of plots (a) and (b) shows that the {\em local} volume fraction of tails of the 
hexagonal-former at the diaphragm rim increases by about 20 \%, 
and that the {\em local} density of 
heads of this amphiphile increases there by almost 50\%.  

The barrier to fusion in the standard mechanism is shown in the upper
curve of 
Fig. \ref{barrier1} as a function of
concentration of the hexagonal-forming amphiphile. 
One sees that the dependence is non-linear. The effect of the
hexagonal-forming amphiphile in reducing the barrier to fusion is
greatest when this amphiphile is first added, as it can go to the region
where it relieves the most strain. As more and more is added, its
ability to reduce the barrier to fusion is lessened.

The barrier to fusion in the first stalk-hole mechanism is shown in the lower
curve. We have assumed a reasonable energy of 4 $k_BT$
for the defects that appear at the end of the elongated stalk which
partially surrounds a hole in one of the bilayers.
That the barrier to fusion is somewhat lower in the first stalk-hole mechanism
than in the standard one, and is much less sensitive to the architecture
than is the standard mechanism
for a system composed primarily of amphiphile characterized by $f=0.4$
could have been anticipated by the results presented in Fig. 10 of
reference \cite{Katsov06}. 
As 
seen there, for $f=0.35$, and $\gamma/\gamma_0=0.2,$ the 
barrier to fusion is somewhat lower in the first stalk-hole mechanism than in 
the standard mechanism, and the barrier in the latter varies more 
rapidly with architecture, $f$, than in the stalk-hole mechanism. 

In the upper panel of Fig. \ref{barrieralta}, 
we compare the fusion barriers in the first and
second stalk mechanisms. The former is shown in filled circles and the
latter is shown in filled triangles. Defect energies are taken to be
$4\ k_BT$. One sees that there is not a
great deal of difference in the energy barriers in the two mechanisms.
One also notes that the second stalk-hole mechanism has a lower energy
than that of the first when the fraction of hexagonal-forming
amphiphiles is low. The situation is reversed as the fraction
increases. This is to be expected as the second stalk-hole
intermediate consists of portions of a fusion pore and of two
holes. Both of these structures are disfavored by the
hexagonal-forming amphiphiles.
On the other hand, the first stalk-hole intermediate
consists of portions of a hemifusion diaphragm, and only one hole. The
hemifusion diaphragm is favored by the hexagonal-forming amphiphiles.

The lower panel of  Fig. \ref{barrieralta} illustrates that the barrier
to fusion in the stalk-hole mechanism is not very sensitive to the
choice of defect energy. The barriers heights are shown there for the
first stalk-hole mechanism for the case in which the defect energy is $4
k_BT$ (open circles) and in which the defect energy vanishes (closed
circles).

\section{Results for asymmetric bilayers}

We now consider the situation in which the compositions of the
two different leaves of the bilayer differ. In particular, we will
fix the composition of the hexagonal-forming lipid in the cis leaf. The
overall composition of lamellae- and hexagonal- forming lipids in the bilayer is still
controlled by the activities $z_1,\ z_2$ and $z_s$ and the incompressibility
condition. Therefore we want to calculate the excess free energy 
$\delta\Omega^{asym}(T,A,z_1, z_2,z_s,n_2^{cis})$, where 
$n_2^{cis}$ is the number
of hexagonal-forming lipids in the cis leaf of the bilayer;
\begin{eqnarray}
n_2^{cis}&=&\frac{1}{\alpha N v}\int d{\bf r}\ \phi_{2}({\bf
r}),\nonumber \\
 &=&\frac{1}{\alpha N vf_2}\int d{\bf r}\ \phi_{A,2}({\bf r}).        
\end{eqnarray}
The integral is over the volume of the cis leaf of the bilayer. 
In the second line, we determine the number of hexagonal-forming lipids
in the cis layer by counting the number of their head groups, which will
be more convenient.
Rather than calculate the free energy in an ensemble in which 
the number of hexagonal-forming lipid heads is fixed,   it is
far easier, as usual,  
to calculate the free energy in an ensemble in which a local
field, $h({\bf r})$,
controls the average local average value of $\phi_{A,2}({\bf r}),$ 
and therefore of $n_2^{cis}.$ 
This adds to the system's internal energy a term of the form
\begin{equation}
-\frac{k_BT}{Nv}\int\ d{\bf r}h({\bf r})\phi_{A,2}({\bf r})
\end{equation}
The field $h({\bf r})$ is taken to be non-zero only in the cis leaf, and
will be discussed further below. By a simple extension of the procedure
for the  symmetric bilayer we obtain for the excess free energy in this
ensemble, 
$\delta{\tilde\Omega}^{asym}(T,A,z_1,z_2,z_s,[h])$,
\begin{eqnarray}
\label{free2}
\frac{Nv}{k_BT}\delta{\tilde\Omega}^{asym}&=&-z_1Q_1(T,[w_A,w_B])
-z_2Q_2(T,[w_A-h,w_B])-z_sQ_s(T,[w_A])\nonumber \\
                  &&+\int d{\bf r}[\chi N\phi_A({\bf r})\phi_B({\bf r})-
w_A({\bf r})\phi_A({\bf r})-w_B({\bf r})\phi_B({\bf r})\nonumber \\
                  &&-\xi({\bf r})(1-\phi_A({\bf r})-\phi_B({\bf r}))].
\end{eqnarray} 
The self-consistent equations, \ref{sceq1} to \ref{sceq5}, are
unaffected. Again the free energy in the self-consistent field
approximation is obtained by substituting the functions which satisfy
the self-consistent equations 
into the free energy of Eq. \ref{free2} with the result
 \begin{eqnarray}
 \label{scasymfree1}
  \frac{Nv}{k_BT}\delta{\tilde\Omega}^{asym}_{scf}&=&
-z_1Q_1(T,[w_A,w_B])-z_2Q_2(T,[w_A-h,w_B])-z_sQ_s(T,[w_A])\nonumber\\
  &-&\int\ d{\bf r}\chi N\phi_A({\bf r})\phi_B({\bf r}),
\end{eqnarray}
The desired free energy, 
$\delta\Omega^{asym}_{scf}(T,A,z_1,z_2,z_s,n_2^{cis})$, is now obtained by a
Legendre transform
\begin{eqnarray}
\label{free3}
\frac{Nv}{k_BT}\delta\Omega^{asym}_{scf}(T,A,z_1,z_2,z_s,n_2^{cis})&=&\frac{Nv}{k_BT}\delta{\tilde\Omega}^{asym}_{scf}(T,A,z_1,z_2,z_s,[h])\nonumber
\\
&+&\int\ d{\bf r}h({\bf r})\phi_{A,2}({\bf r})
\end{eqnarray}
so that
\begin{eqnarray}
\frac{Nv}{k_BT}\delta\Omega^{asym}_{scf}
&=& -z_1Q_1(T,[w_A,w_B])-z_2Q_2(T,[w_A-h,w_B])
 - z_sQ_s(T,[w_A])\nonumber\\
&+&\int\
d{\bf r}[h({\bf r})\phi_{A,2}({\bf r})-\chi N\phi_A({\bf r})\phi_B({\bf r})].
\end{eqnarray}
Because the system is constrained to have a different concentration of
hexagonal-formers in the cis leaf than in the trans leaf, its free
energy will clearly be greater than if it were not so constrained. This
is also true of the free energies of the various intermediates, like the
stalk, hemifusion diaphragm, and pore. For the fusion process, however,
we are interested in differences in free energies between the
intermediates and the flat bilayers, and these differences can certainly
be less in the constrained system. 
\subsection{Standard Mechanism}
The calculations for the standard mechanism are relatively
straightforward due to the axial and reflection symmetry of the stalk,
the hemifusion diaphragm, and the pore. We need only indicate where the
field $h({\bf r})$ is non-zero, which is shown in Fig. \ref{fieldlocation}. 
Specifically 
\begin{eqnarray}
h({\bf r})=h(z,\rho)&=&h_0\qquad {|z|\leq0.6 R_g\ {\rm and}\ \rho\geq
R+0.6R_g},\nonumber \\
         &=& 0 \qquad{\rm otherwise},
\end{eqnarray} 
with $R$ the radius of the hemifusion diaphragm
defined previously \cite{Katsov04}.  

In Fig. \ref{asymfexcess} we show results for a bilayer under a tension $\gamma/\gamma_0=0.2$ composed of the lamellae-former which comprises a
fraction $\phi_1=0.650$ of
the bilayer by volume, and the hexagonal-former  comprising a fraction $\phi_2=0.350$ by volume.  
Results are presented for the excess free energy of the
hemifusion diaphragm (solid lines) and of the fusion pores (dashed
lines) for different volume fractions in the cis leaf of the hexagonal-forming
amphiphile. In the upper set of curves, there is no asymmetry, so that the
volume fraction of hexagonal former in the cis leaf,
$\phi_2^{cis}=0.350$,  is the same as in the whole bilayer. In the middle
curve, the volume fraction of the hexagonal former in the cis leaf,
has been increased to $\phi_2^{cis}=0.395$. Its volume fraction in the
trans leaf is concomitantly reduced to $\phi_2^{trans}=0.305$, and the volume fractions of
the lamellae-former in the cis and trans leaves are 0.605 and 0.695,
respectively. In the lowest curve, we have set $\phi_2^{cis}=0.431$, so
that $\phi_2^{trans}=0.269$, and the volume fractions of the
lamellae-former in the cis and trans leaves is 0.569 and 0.731. The
barrier to fusion is reduced from $11k_{B}T$ to $8.5k_{B}T$, 
to $5k_BT$ as the asymmetry increases.  For the largest asymmetry shown,
the barrier
to fusion is essentially no greater than the barrier to formation of the
initial stalk itself.
\subsection{Stalk/hole mechanism}
 We have calculated the barrier to fusion between asymmetric bilayers in
the second stalk-hole mechanism. We have chosen this path, rather than
the first stalk-hole mechanism, because the latter involves the calculation of
the free energy of a hole in an asymmetric bilayer and of a hemifusion
diaphragm which consists of the cis and trans layer of one of the
original bilayers. As the bilayer is not symmetric, neither is the
hemifusion diaphragm, and this lack of symmetry about the $x,y$ plane
makes the calculation rather slow. The second stalk-hole mechanism does not
involve this asymmetric hemifusion diaphragm, although it still involves
holes in asymmetric bilayers. The calculation of the energies in this
pathway is more rapid. We have already shown that there is not a
great deal of difference in the barrier energies in the two pathways
in symmetric bilayers,
Fig. \ref{barrieralta}(a), and assume the same is true with asymmetric 
bilayers. If anything, we will overestimate the fusion barrier of the
stalk-hole mechanism because,
as we add
hexagonal-formers, the barrier in the second stalk-hole pathway which 
we calculate will probably become somewhat larger than that in the 
first pathway,
just as it is in the symmetric bilayer case, Fig. \ref{barrieralta}(a).

Our results for the barrier to fusion of asymmetric bilayers 
within the second stalk-hole mechanism are shown in
Fig. \ref{comparison}. We have calculated them for bilayers in which the
average volume fraction of hexagonal-formers in the entire bilayer is
kept fixed at $\phi_2=0.350$, while the fraction of hexagonal-formers in
the cis layer, $\phi_2^{cis}$, takes the values $\phi_2^{cis}=0.350$
(i.e. no asymmetry), $\phi_2^{cis}=0.395$, and $\phi_2^{cis}=0.431.$
The barrier to fusion for this second stalk-hole pathway is shown by the
triangles. The values of $\alpha$ at the saddle point in the fusion
pathway are, $\alpha=0.073$ for $\phi_2^{cis}=0.350$, 
$\alpha=0.174$ when $\phi_2^{cis}=0.395$, and $\alpha=0.18$ when $\phi_2^{cis}=0.431$. 
These barriers to fusion are compared to those 
calculated in the standard mechanism and shown in squares. These values 
were shown previously in Fig.\ref{asymfexcess}. Finally, we also compare
them with the free energies of the stalk, shown in circles.

We note that the small values of $\alpha$ in the stalk-hole mechanism 
imply that the stalk does not have to elongate very much in order to 
nucleate the formation of the two holes which, when surrounded by the 
stalk, will become the fusion pore. (We recall that in surrounding the 
holes, the energy of the system is reduced as the line tensions of the 
bare holes are replaced by the lower line tension of a hole next to a 
stalk \cite{Katsov06}.)  Hole formation is enhanced, and the barrier to 
fusion reduced, because the majority amphiphile, $\phi_1=0.65$, is a 
lamellae-former with $f=0.4$. Furthermore the actual volume fraction of 
the lamellae-former near the rim of a hole will be larger than this 
because the amphiphiles are free to move within a leaf to that region 
where they will reduce the energy most. The increase of $\alpha$ with the 
fraction of hexagonal-formers in the cis leaf is readily understood. As 
the fraction of hexagonal-former in the cis leaf increases, the energy of 
a pore decreases, as noted previously. It follows from Eq. 
\ref{freeafter2} that $\alpha$, the fraction of the stalk-hole 
intermediate that resembles a pore, will increase.

\section{Discussion}
We have employed a model of a mixture of two amphiphiles, one which is a
lamellae-former, the other a hexagonal-former. The ratio of their
hydrophilic part to the entire molecule was chosen so that the first
resembles DOPC and the latter resembles DOPE. The two have the same
hydrophobic volumes, but different hydrophilic ones.
We have solved the
model within self-consistent field theory. 

We first considered bilayers
whose leaves have identical compositions, and added hexagonal-formers to
each leaf equally. We examined the effect of this addition 
on the
barrier to fusion as calculated in the standard mechanism, and the first
and second stalk-hole mechanisms. The barrier energy is reduced significantly
in the standard mechanism from about $24\ k_BT$ with no
hexagonal-formers to about $11\ k_BT$ with a volume fraction of 0.35
hexagonal-formers. This is seen in Figs. \ref{barrier1} and
\ref{comparison}. 
As noted earlier, \cite{Katsov04}, we 
expect that the energies in biological, lipid, membranes are higher by a 
factor of about 2.5 than in the block copolymer membranes which 
we consider. 
Thus the above barrier values would correspond to one of $60\ k_BT$
being reduced to $28\ k_BT$. The reduction in the fusion barrier of the
standard mechanism is due to a reduction in the energy of the hemifusion
intermediate, partly because the average number of hexagonal-formers has
increased \cite{Kozlovsky02}, and partly because
the hexagonal-forming amphiphiles preferentially go to the edge of the 
hemifusion diaphragm, as seen in Fig. \ref{nonuniform}.
Fig. \ref{barrier1}  shows that the greatest rate of
decrease comes about when the hexagonal-formers are first added to the
pure bilayer of lamellae-formers. This rapid decrease occurs because the
hexagonal-formers go to the regions where they
can most readily reduce the free energy. The distribution of the
different amphiphiles are not spatially uniform when there are fusion 
intermediates.  
The reduction of the barrier energy in the stalk-hole mechanism, is more
modest but not insignificant. In the second stalk-hole mechanism, 
it is reduced from
$8.3\  k_BT$ when there are no hexagonal-formers to $6.8 \ k_BT$
with a 
volume fraction of 0.350 hexagonal-formers. Again this would correspond
to a reduction from $21\ k_BT$ to $17\ k_BT$.
  
We then examined the effect on the barrier to fusion of an unequal 
distribution of 
hexagonal- and lamellae-formers in the two leaves. We considered a
system in which the
hexagonal-formers make up a volume fraction of  0.350 of the whole
system, much as they do in 
human red blood cell membranes. This brings about a further significant 
reduction 
in the barrier to fusion in both mechanisms. In the standard mechanism, 
this is  due to the reduction in
energy of the hemifusion diaphragm, while in the
stalk-hole mechanism it is due primarily to the reduction in energy of 
the elongated stalk. 

The energy of the initial stalk itself is not affected very much either by the
addition of hexagonal-formers to each leaf equally, as seen in
Fig. \ref{oldmech}, nor by the redistribution of the hexagonal-formers
between the two leaves, Fig. \ref{comparison}. The former result is in
contrast to the prediction of phenomenological theories of a sensitive
dependence upon the amount of hexagonal-formers, \cite{Kozlovsky02}.

Certainly the most important result of our calculation is the following:
although 
the fusion process remains one with two barriers, one due to stalk formation and 
another which depends upon the specific mechanism, 
the second barrier is rapidly reduced by the addition of
hexagonal-former of greater abundance in the cis layer to a 
value comparable to that of the initial stalk itself.  
As emphasized earlier, the calculated energy of the 
stalk is rather small, on the order of $5\ k_BT$ in our copolymer
system, corresponding to $13\ k_BT$ in a biological 
membrane. 

We note that the volume fraction of hexagonal-former in 
the cis leaf at which the two barriers become approximately equal occurs
in our model at a value of about $\phi_2^{cis}\sim 
0.43$. The average fraction of hexagonal-formers in the bilayer is
0.35. Assuming equal molecular weights for the A and B components of the
diblock, these volume fractions correspond to a mole fraction of 0.47 in
the cis leaf of a bilayer in which its average mole fraction is 0.39. 
Again, the mole fractions of hexagonal-formers in the membrane of human red
blood cells is approximately 0.54 in the cis leaf and 0.35 when averaged
over both leaves of the bilayer. Thus equality of the two barriers
occurs in our model at a somewhat smaller asymmetry between leaves than occurs in
red blood cell membranes. As the asymmetry increases, the second barrier
to fusion continues to decrease and eventually becomes negative. When
this occurs in a bilayer under zero surface tension, the bilayer is unstable.
In the system shown in Fig. \ref{comparison}, this instability occurs at
a mole fraction of hexagonal-former in the cis layer of about 0.50.
   
To reiterate, our major result is that the two barriers to fusion 
are {\em comparable} and {\em small} for an
amount of hexagonal-former found in the cis layer which does not differ
greatly from that found in red blood cell membranes. If this result  
is applicable to biological membranes, then there are 
important implications.  The first is that fusion should proceed readily
once external sources have 
brought the membranes sufficiently close to initiate the process. 
This shifts the focus of fusion to an understanding of those mechanisms which 
bring this about.  
The second is an emphasis on the caution which must be exercised when
extrapolating to the fusion of biological membranes the 
experimental or theoretical
results gleaned from fusion studies of non-biological membranes with 
leaves of equal compositions. 


We are grateful to Kirill Katsov for useful correspondence. This work
was supported by the National Science Foundation under grant No. 0503752.

\clearpage
\clearpage
\clearpage
\begin{figure}
   \begin{center}
       \includegraphics*[width=5.25in]{fexcessoldmech.eps}
      \caption{}
      \label{oldmech}
   \end{center}
\end{figure}
\clearpage
\begin{figure}
   \begin{center}
       \includegraphics*[width=5.25in]{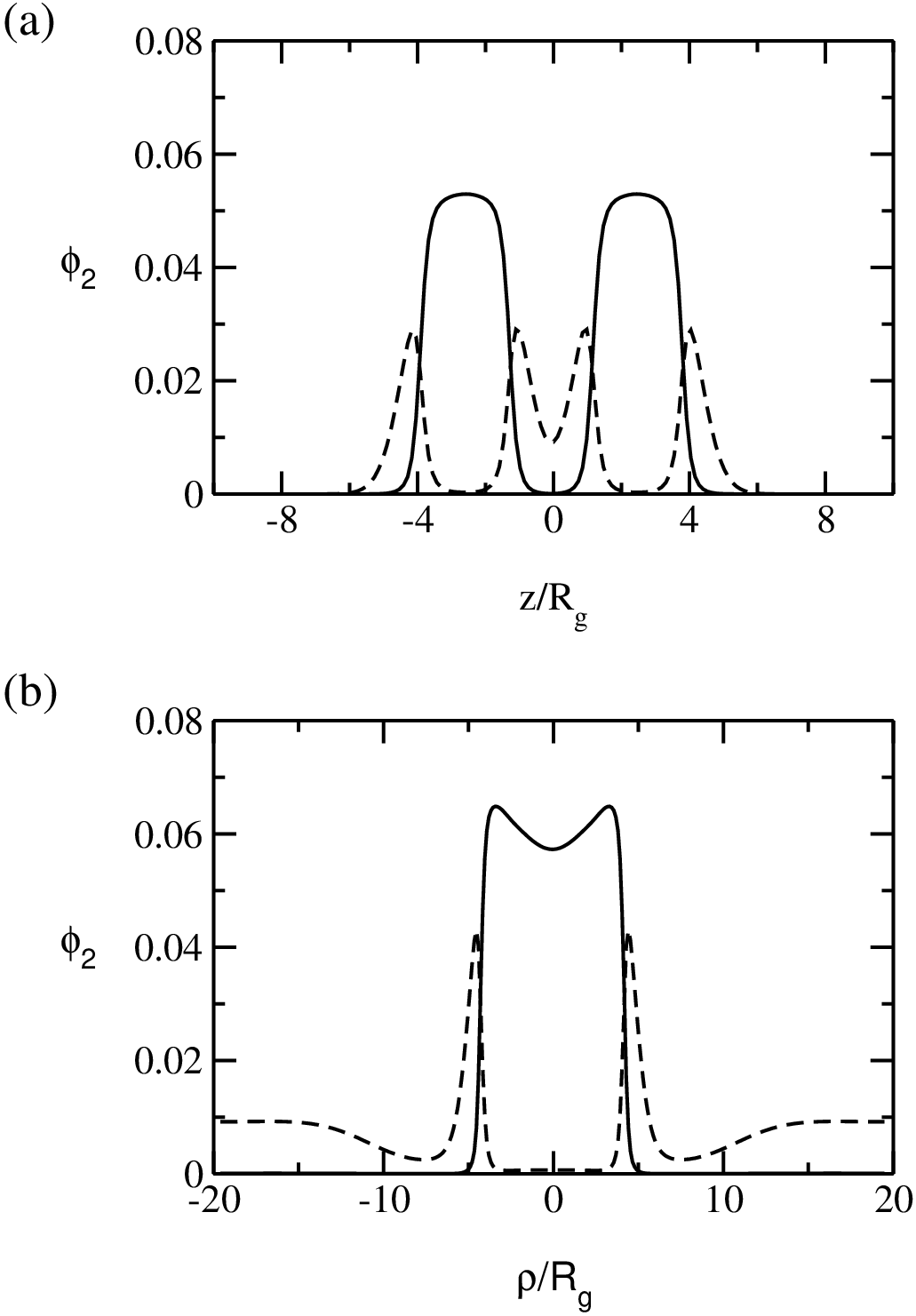}
      \caption{}
      \label{nonuniform}
   \end{center}
\end{figure}
\clearpage
\begin{figure}
   \begin{center}
   \includegraphics*[width=5.25in]{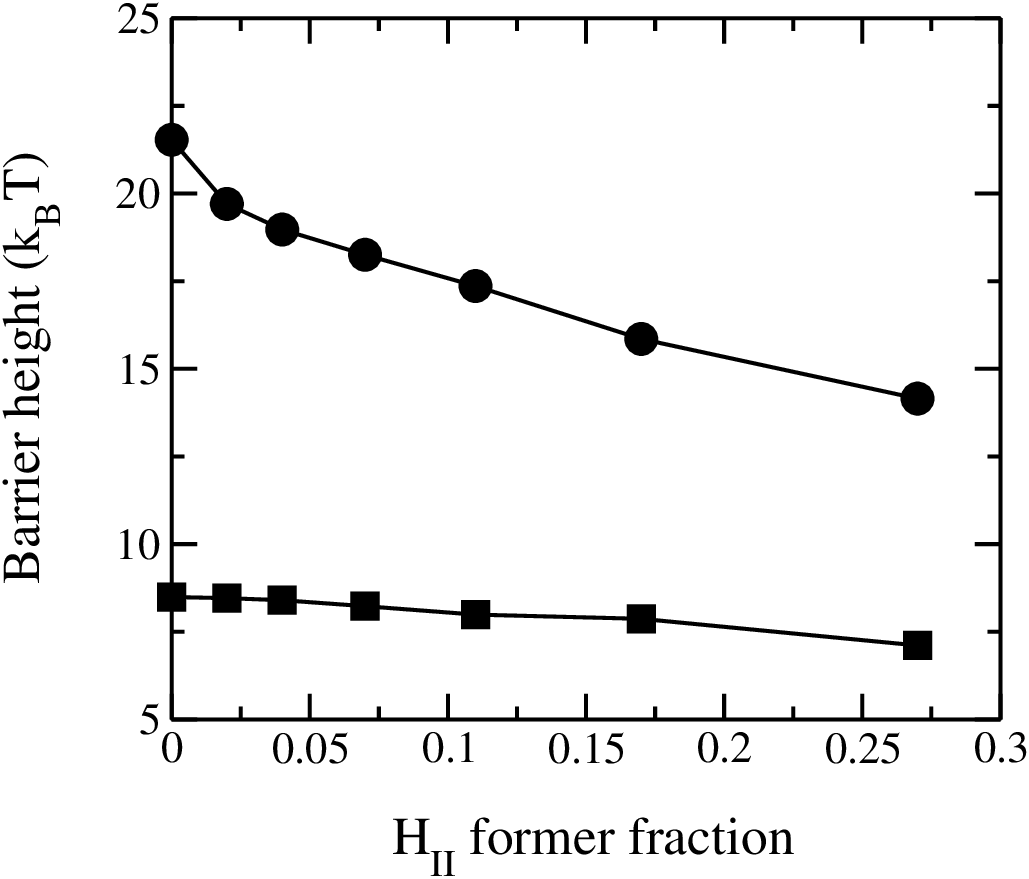}
      \caption{}
      \label{barrier1}
   \end{center}
\end{figure}
\clearpage
\begin{figure}
   \begin{center}
   \includegraphics*[width=5.25in]{barrier2.eps}
      \caption{}
      \label{barrieralta}
   \end{center}
\end{figure}
\clearpage
\begin{figure}
   \begin{center}
      \includegraphics*[width=5.25in]{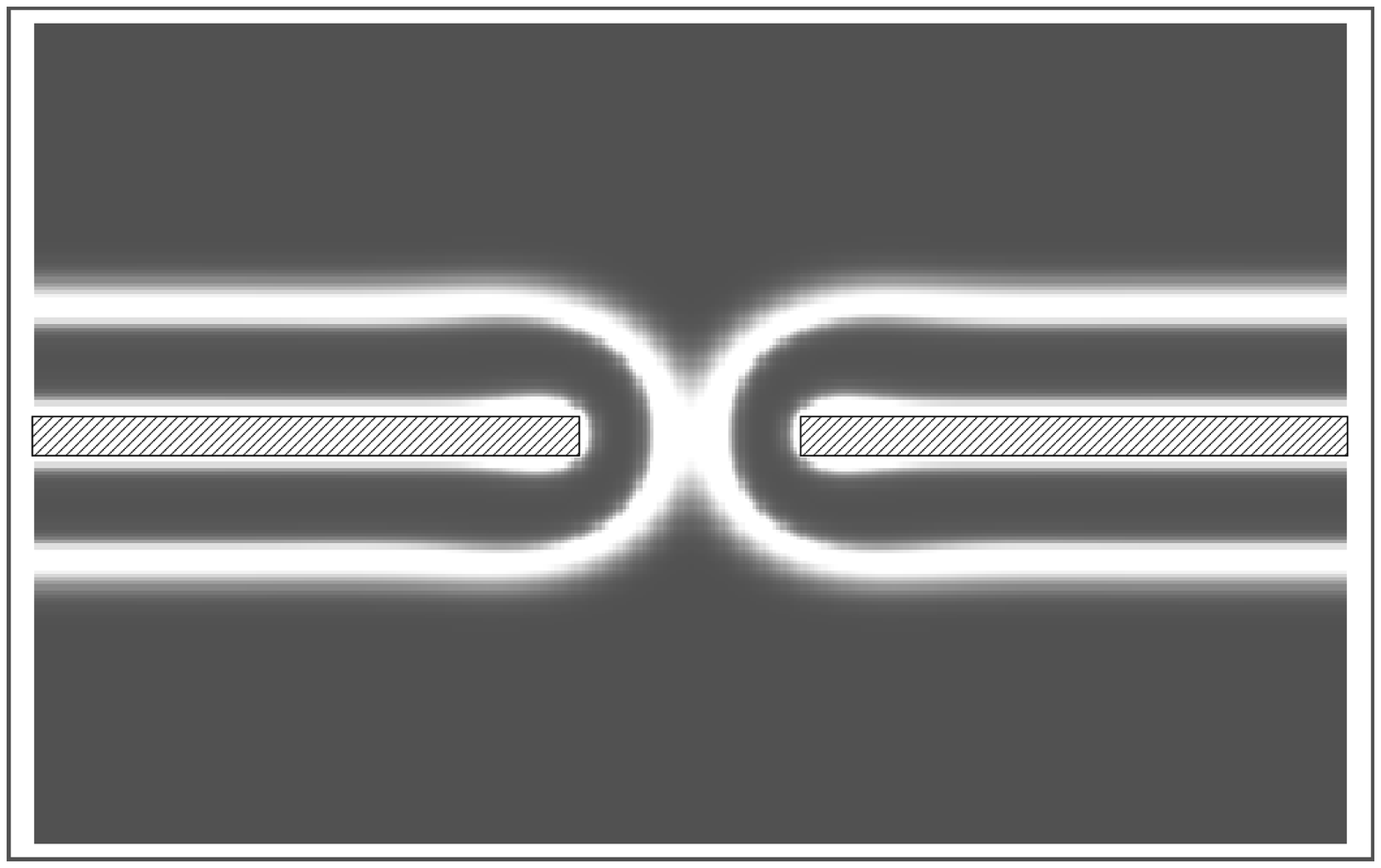}
      \caption{}
      \label{fieldlocation}
   \end{center}
\end{figure}
\clearpage
\begin{figure}
   \begin{center}
       \includegraphics*[width=5.25in]{asymfexcess.eps}
\caption{}
      \label{asymfexcess}
   \end{center}
\end{figure}
\clearpage
\begin{figure}
   \begin{center}
       \includegraphics*[width=5.25in]{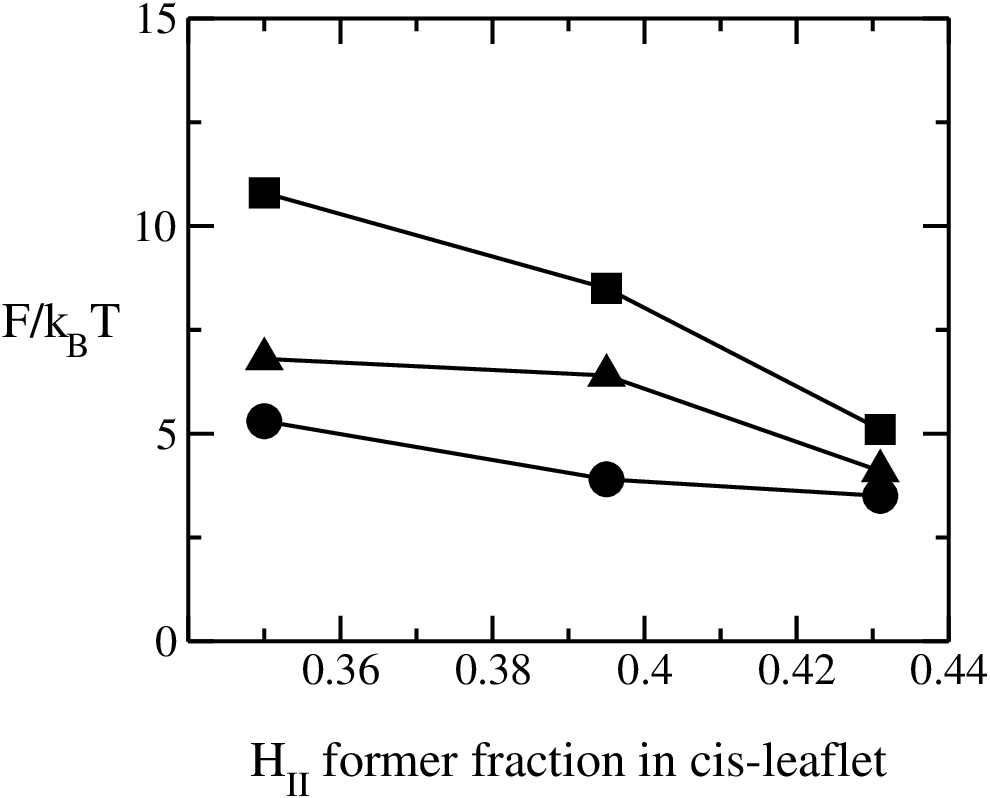}
\caption{}
      \label{comparison}
   \end{center}
\end{figure}
\clearpage
\section{Figure Captions}
\begin{itemize}
\item[]{Figure 1.} The standard stalk model description of membrane 
fusion.  Light regions indicate the areas  of head groups of the bilayer 
in the left-hand panel, and of tail groups in the right-hand panel.  
(a) stalk (b) hemifusion 
diaphragm (c) fusion pore.
\item[]{Figure 2.} (a) Parameterization of the elongated stalk. The 
shading schematically shows the location of the hydrophobic segments in 
the plane of symmetry between fusing bilayers. The arc radius $R$ 
corresponds to the radial distance to the outer hydrophilic/hydrophobic 
interface in the plane of symmetry. Values of the fractional arc angle, 
$\alpha$, defined in the range [0,1], are given at the top of each stalk 
configuration. Note that $\alpha=0$ corresponds to the original stalk 
configuration. (b) Parameterization of the stalk-hole complex. In the 
first stalk-hole mechanism, there is a hole in one bilayer and the 
projection of its edge is shown with a dashed line. In the second 
stalk-hole mechanism, there is a hole in each of the bilayers, and the 
dashed line represents the projection of their edges. The radius of the 
hole, or holes, is $R-\delta$. The hydrophobic thickness of the bilayer is 
$\delta$.  Values of the fractional arc angle, $\alpha$, defined in the 
range [0,1], are given at the top of each configuration.
\item[]{Figure 3.} Excess free energies of fusion intermediates in the 
standard model are shown at a  tension of $\gamma/\gamma_{0}=0.2$.  
Solid curves indicate stalk/hemifusion intermediates and dashed curves 
fusion pores.  The bilayers consist of AB diblocks of two different 
lengths and architectures.  The first diblock is described by $N$ segments and
$f_{1}=0.4$, and the second diblock by ${\tilde\alpha}N$ segments with
${\tilde\alpha}_{2}=0.85$ and 
${\tilde\alpha}_{2}(1-f_{2})=0.6$.  From top to bottom, the volume fractions of 
type 2 diblocks in the bilayers are 0.00, 0.04, 0.11, and 0.17.
\item[]{Figure 4.} (a) Volume fractions, $\phi_2$, of the head group, dashed line, and
tail, solid line, of the hexagonal-forming amphiphile in the bilayers  far from the 
hemifusion diaphragm are shown in  a cut perpendicular to the bilayers as
a function of the dimensionless vertical coordinate $z/R_g.$ (b) These same volume fractions
 are shown in the $z=0$ plane of symmetry which passes through the hemifusion
 diaphragm itself as a function of the dimensionless radial coordinate
 $\rho/R_g$. The hemifusion diaphragm has a radius of about $5R_g.$ 
\item[]{Figure 5.} Barrier height of fusion process as a function
of the volume fraction of the hexagonal-forming ($H_{II}$) amphiphile. 
The upper curve shows the 
barrier heights in the standard stalk-hemifusion mechanism. The lower 
curve shows the barrier heights in the first stalk-hole mechanism with 
a defect energy of $F_d=4k_{B}T$.
\item[]{Figure 6.}(a) Comparison of the barrier to fusion in the first 
(filled circles) and second (filled triangles) stalk-hole mechanisms as a 
function of volume fraction of hexagonal-forming ($H_{II}$) amphiphile. (b) 
Comparison of barrier to fusion in the first stalk-hole mechanism with 
defect energy of $4 k_BT$ (open circles) and vanishing defect energy 
(closed circles).
\item[]{Figure 7.} Density profile of a fusion pore. The region where 
external fields are applied to maintain asymmetry is marked by shaded 
areas on the density plot of small head groups.  White regions indicate the 
areas where small head groups are concentrated and the gray regions the 
areas in which their concentration is strongly reduced.
\item[]{Figure 8.} Excess free energies of standard fusion intermediates 
for bilayers of the same overall composition, but with varying 
transbilayer distributions under $\gamma/\gamma_0=0.2$ tension.  The 
bilayers here contain 65\% lamellae-forming diblock  and 35\% 
hexagonal-forming diblock. 
The solid 
curves represent excess free energies of stalk/hemifusion diaphragm and 
the dashed curves excess free energies of fusion pores.  In the upper set 
of curves, there is no asymmetry, so that the volume fraction of hexagonal 
former in the cis leaf, $\phi_2^{cis}=0.350$, is the same as in the whole 
bilayer. In the middle curve, the volume fraction of the hexagonal former 
in the cis leaf, has been increased to $\phi_2^{cis}=0.395$. In the lowest 
curve, we have set $\phi_2^{cis}=0.431$. The barrier to fusion is reduced 
from $11k_{B}T$ to $8.5k_{B}T$, to $5k_BT$ as the asymmetry increases.
\item[]{Figure 9.} Comparison of the barrier to fusion of asymmetric 
bilayers containing an average volume fraction of hexagonal-formers of 
$\phi_2=0.35$ as calculated along the standard pathway, (squares), and the 
second stalk-hole pathway, (triangles) for three different volume 
fractions of hexagonal-formers in the cis layer; $\phi_2^{cis}=0.350, \ 
0.395, \ 0.431$. Also shown is the free energy of a stalk, (circles), in 
the same systems.
\end{itemize}
\clearpage
\bibliography{kkfusion06}
\end{document}